# Sparse Multipath Channel Estimation using DS Algorithm in Wideband Communication Systems

Guan Gui[1,2], An-min Huang[1], Qun Wan[1]

[1] Dept of Electric Engineering, University of electrical Science and Technology of China, Chengdu, China
[2] Dept of Electrical and Communication Engineering, Graduate School of Engineering, Tohoku University, Sendai, Japan.

*Abstract*-Wideband wireless channel is a time dispersive channel and becomes strongly frequency-selective. However, in most cases, the channel is composed of a few dominant taps and a large part of taps is approximately zero or zero. They are often called sparse multi-path channels (MPC). Conventional linear MPC methods, such as the least squares (LS), do not exploit the sparsity of MPC. In general, accurate sparse MPC estimator can be obtained by solving a LASSO problem even in the presence of noise. In this paper, a novel CS-based sparse MPC method by using Dantzig selector (DS) [1] is introduced. This method exploits a channel's sparsity to reduce the number of training sequence and, hence, increase spectral efficiency when compared to existed methods with computer simulations.

## I. INTRODUCTION

With the number of wireless subscribers increasing daily, and new wireless initiatives generating massive data traffic on the rise, the already greedy demand for high-speed data services is only getting more insatiable. Consequently, how to overcome the scarcity of spectral resources to meet the ever-growing need for high data rate is going to be a continuous challenge for communication engineers. One way to achieve a high data rate is to simply increase the transmission speed. Since the multipath channels are widely separated in time, creating a large delay spread which results in sampled channel easily reaches hundreds of symbol intervals. As the delay spread sampled length becoming longer, consequently, accurate channel estimation requires redundant training sequences or pilots.

Thus, this procedure results in spectral resources waste. Thanks to many near zero-valued taps in a sampled channel, therefore, accurate knowledge of the underlying multipath channel often depends on the number of large-magnitude taps instead of the sampled channel length (Fig. 1), where channel length is $p$=80 and the number of dominant taps is $S$=4 which relative to the channel length is small. Thus, this sampled channel can be called sparse multipath channel. Several methods have been proposed for sparse multiptah channel estimation, such as matching pursuit (MP) [2], orthogonal matching pursuit (OMP) [3], Lasso [4, 5]. In general, these methods can be concluded two types: greedy algorithm and convex program. However, the greedy algorithm is less stable than convex program which is hard to implement and complex to compute in real channel estimation in the presence of noise.

Based on the recent development of compressed sensing [6] or compressive sampling [7], which is a non-adaptive compressive method that takes advantage of natural sparsity at the input and is fast gaining relevance to both researchers and engineers for its universality and applicability, we have a new perspective for channel estimation. In this paper, we focus on a novel sparse channel estimate method known as Dantzig selector (DS) [1], which uses $L_1$-minimization with regularization on the maximal estimation residuals. Even if the number of variables is much larger than the number of training sequence, DS can be acquired the accurate solution relative to the above mentioned methods. In other words, if we demand same channel estimation performance, DS method needs shorter training sequence. Thus, using this method can economize the spectral resources and increase spectral efficiency in wideband communication systems.

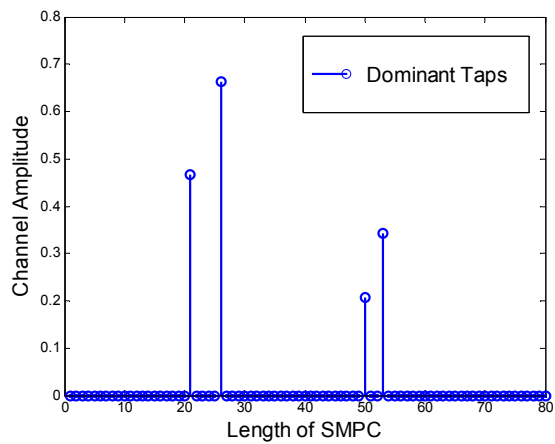

Figure 1. An example of sparse multipath channel (SMPC). This is a simply sparse channel model which compounds the propagated amplitude fading and the delay spread.

The nice concept of DS was proposed by Candes and Tao in 2007. They establish optimal mean square error rate properties under a sparsity scenario, when the number of nonzero taps of the true sparse channel of parameters is small. Thus, the method can obtain approximate sparsest solution which close in oracle channel estimator, which can be considered as the lower bound on the MSE of estimator in sparse channel estimation. It is worth noting that the unknown complex channel vector to be estimated is deterministic. Thus, oracle estimator can be obtained if we know the position of dominant channel taps. Quite recently, some mathematicians [1], [8] have discussed the relationship between Dantzig selector and Lasso estimator. Peter J. Bickel, Y. Ritov and A. Tsybakov [8] have exhibited an approximate equivalence between the Lasso estimator and DS. While T. Tony Cai and Lv presented a promising fact that the Dantzig selector solves a linear program usually faster than the existing methods such as Lasso [9]. In the end of the paper, we will also compare channel estimation performance between DS and Lasso. Although no provable results have been attained, empirical studies have suggested the DS method slight outperform the Lasso.

## II. SPARSE MULTIPATH CHANNEL MODEL

At first, the symbols used in this paper are described as following. The superscript $^H$ stands for Hermite transposition. Bolded capital letters denote a matrix where bolded lowercase letters represent a vector. Notation $|\cdot|$ stands for the absolute value. Norm operator $\|\cdot\|_0$ denotes $L_0$ vector norm, i.e., the number of non-zero entries of the vector; $\|\cdot\|_1$ denotes $L_1$ vector norm, which is the sum of the absolute values of the vector entries. $\|\cdot\|_2$ denotes $L_2$ norm. $\tilde{\mathbf{h}}$ and $\mathbf{h}$ indicate estimate channel vector and actual channel vector, respectively.

We consider single-antenna wideband propagation systems, which are often equivalent to frequency-selective baseband channel model. The equivalent baseband transmitted $\mathbf{X}$ and received signals $\mathbf{y}$ is given by

$$\mathbf{y} = \mathbf{X}\mathbf{h} + \mathbf{z} \tag{1}$$

where $\mathbf{X}$ is a complex training signal with Toeplitz structure of $N \times L$ dimensions. $\mathbf{z}$ is the $N \times 1$ complex additive white Gaussian noise (AWGN) with zero mean and variance $\sigma^2$. $\mathbf{h}$ is an $L \times 1$ unknown deterministic channel vector which is given by

$$\mathbf{h}(\tau) = \sum_{i=0}^{L-1} h_i \delta(\tau - \tau_i), \tag{2}$$

where $h_i$ are complex channel coefficients and $\tau$ is a delay spread which sampling length $L$ in baseband channel representation. $T = \#\{|h_i| > 0, i \in L\}$ denotes the number of dominant taps of the SMPC where $T \ll L$. Suppose that there are $T$ dominant channel taps distributed randomly over the channel.

## III. COMPRESSED SENSING OVERVIEW

Compressed sensing (CS)-based sparse reconstruction or approximate, one aims to approximate or reconstruct a sparse or compressible signal with only a limited number of linear measurements. Using a sensing training sequence and having knowledge of these measurements, the sparsest channel vector giving rise to these measurements is sought. However, the sparsest solution ($L_0$-regularization) is always a Non-deterministic Polynomial-time hard (NP-hard) problem. Donoho [10] has presented a necessary and sufficient condition (RIP) on the sensing matrix (training sequence) so that every channel vector is a point of $L_1/L_0$ -equivalence. In this context, Candès and Tao [11] introduced the restricted isometry constants (RIC) of a matrix, which is called training sensing sequence in this paper. The $T$-RIC of a $N \times L$ training sequence $\mathbf{X}$, denoted by $\delta_T$, is defined as the smallest value $\delta_T$ ($\delta_T \in (0,1)$) which can satisfy the inequality

$$(1-\delta_T)\|\mathbf{h}\|_2^2 \leq \|\mathbf{X}\mathbf{h}\|_2^2 \leq (1+\delta_T)\|\mathbf{h}\|_2^2 \tag{3}$$

for any SMPC vector $\mathbf{h}$. If (3) is satisfied, the training sequence $\mathbf{X}$ is said to satisfy RIP of order $T$ and accurate channel estimator with high probability can be obtained by using CS methods. From CS perspective, research on the RIP of the training sequence has two important purposes. First, RIP-based training sequence is a sufficient condition to robust probe sparse channel dominant taps. Furthermore, in the process of error performance analysis, RIC of training sequences play important role to improve lower bound of MSE performance.

## IV. CHANNEL ESTIMATION USING DS ALGORITHM

To estimate $\mathbf{h}$ with the observed vector which is corrupted by noise, we can use a new estimator, namely, Dantzig selector [1], to the convex program problem. In this section, DS algorithm and structure sensing training sequence design method will be introduced.

### A. Design Sensing Training Sequence

Designing a sensing training sequence of appropriate structure is a crucial procedure in sparse channel estimation. We draw the matrix elements $\mathbf{X}_{i,j}$ as independent and identically distributed (i.i.d) random variables. If the measure number of the sensing matrix obeys [6]

$$n \geq c \cdot T \cdot \log(p/T) \tag{4}$$

where $c$ is a small constant, then sensing matrix $\mathbf{X}$ can be shown to have the RIP with high probability. In fact, a strengthened version of the results first reported in Toeplitz-structured matrices [12] satisfies RIP. Based on the linear channel model (1), we compare estimation performance of LS and DS using Toeplitz sensing sequence.

### B. DS Algorithm

Given the observations in (1), the Dantzig selector $\mathbf{h}$ is the solution to the following optimization

$$\min_{\tilde{\mathbf{h}} \in \mathbb{R}^p} \|\tilde{\mathbf{h}}\|_1 \quad \text{subject to} \quad \|\mathbf{X}^H(\mathbf{y} - \mathbf{X}\tilde{\mathbf{h}})\|_\infty \leq \lambda_{DS} \tag{5}$$

where $\lambda_{DS}$ is a constraint parameter which decided by noise and channel environment. The DS ensures that the residual $(\mathbf{y} - \mathbf{X}\tilde{\mathbf{h}})$ is not too correlated with any column of $\mathbf{X}$, which means that all of the significant components in the solution have been accounted for. Accurately, based on the structure of training signal $\mathbf{X}$, we can also design a alternative sensing training signal $\mathbf{X}_{Alt}$ which can further improve MSE performance of sensing DS (SDS) estimator. The basic sensing steps can be written as follows:

a) Correlated residual $\mathbf{W} = diag[|\mathbf{X}^H(\mathbf{y} - \mathbf{X}\tilde{\mathbf{h}})|]$
b) Generate weight factor: $\mathbf{R} = \mathbf{X}\mathbf{W}^2\mathbf{X}^H$
c) Design sensing sequence: $\mathbf{X}_{Alt} = \mathbf{R}^{-1}\mathbf{X}/\mathbf{X}^H\mathbf{R}^{-1}\mathbf{X}$

According to the above training design (*a-c*), more accurate estimator given by

$$\min_{\tilde{\mathbf{h}} \in \mathbb{R}^p} \|\tilde{\mathbf{h}}\|_1 \quad \text{subject to} \quad \|\mathbf{X}_{Alt}^H(\mathbf{y} - \mathbf{X}\tilde{\mathbf{h}})\|_\infty \leq \lambda_{DS} \quad (6)$$

Unfortunately, this estimation method (6) will lead to more compute complexity than DS algorithm in (4). Because of the compute complexity, we consider the DS algorithm for SMPC estimation (4) in this paper. And in future work, we try to simply the compute complexity of sensing DS which is a good candidate algorithm for sparse channel estimation.

An example comparing the sparse reconstruction abilities of LS and DS is illustrated in Figure 2. These methods correspond to using pseudo random sensing training sequence $\mathbf{X} \in \mathbb{C}^{n \times p}$ (where $n = 30$ and $p = 60$) input probe to the sparse channel that has only $T = 5$ nonzero channel taps. The output of the channel is observed at $SNR = 10dB$. As can be seen from Fig.2, the DS estimate is able to identify every nonzero channel taps which are marked with small blue circle. In contrast, it is easy to see that LS method would be unable to identify all the nonzero channel taps which are marked with small red plus.

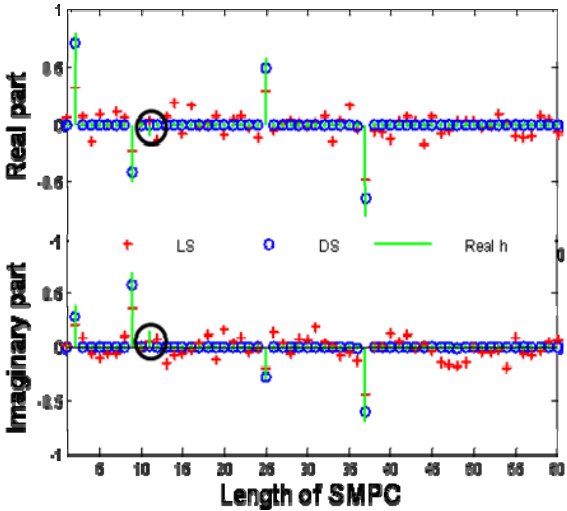

Figure 2. An illustrative example contrasting of LS and DS algorithms for sparse channel estimation.

**Remark:** In the Figure 2, the dominant complex coefficients [0.8+0.4i, -0.5+0.7i, -0.1+0.15i, 0.6-0.3i, -0.8-0.7i] random distributed on overall channel. From a communication theoretic viewpoint, we focus on the non-Bayesian paradigm and assume that both the channel sparsity and the corresponding channel taps are deterministic but unknown. From above figure, we can find that DS method can robust estimate dominant channel taps while neglect the very small taps (such that the small channel coefficients by marked with the black circle). From a communication engineering perspective, small channel coefficients (correspond to channel degree) cannot robust propagate signal and thus it can be consider as noise interference. Thus, DS method considers the coefficients that approximate zero as noise and mitigate it.

*C. Oracle Estimator*

To evaluate the MSE performance of channel estimators, it is very meaningful compare their achievements with theoretical performance bound in practical wideband communication systems, then they are approximate optimal and further improvements in these systems are impossible. This motivates the development of lower bounds on the MSE of estimators on sparse channel estimation. Since the channel vector to be estimated is deterministic, and then we can give a lower bound as for the baseline of MSE. Suppose we know the location set $T = \#\{|h_i| > 0 | i \in T\}$ of dominant channel taps. Thus, the oracle estimator given by

$$\hat{\mathbf{h}}_{oracle} = \begin{cases} (\mathbf{X}_T^H \mathbf{X}_T)^{-1} \mathbf{X}_T^H \mathbf{y}_T, & T \\ 0, & elsewhere \end{cases}, \quad (7)$$

where $\mathbf{X}_T$ is the partial training signal constructed from columns of training signal $\mathbf{X}$ corresponding to the dominant taps of SMPC vector $\mathbf{h}$. It is noting that we call the oracle estimator as oracle bound in Figure 3~4 in the next part.

## V. SIMULATION RESULTS AND DISCUSSION

The parameters used in the simulation are listed in Tab. 1. To illustrate the performance of proposed algorithm, Figure.2 shows the MSE of dominant taps by employing LS, Lasso, DS and oracle estimator. The estimation error using mean square error (MSE) evaluation criterion can be defined as:

$$\text{MSE} \triangleq \text{E}\left\{\|\mathbf{h} - \hat{\mathbf{h}}_m\|_2^2\right\}. \quad (8)$$

TAB. 1 SIMULATION PARAMETERS

| Estimation methods | *Linear algorithm* | LS |
|---|---|---|
| | *Greedy algorithm* | OMP |
| | *Convex program* | Lasso |
| | | DS |
| Channel model | Frequency-selective Rayleigh fading | |
| Channel length *L* | 60 | |
| No. of dominant taps | 4 | |
| Training signal X | Random Toeplitz structure | |
| Monte Carlo | *M*=1000 trails | |

*A. MSE comparison versus SNR*

We compare the performance of the DS and other estimators versus different SNR, which were chosen between 3 and 30. It is seen that, as expected, the performance of all estimators improves with increasing SNR. We again observe large performance gains of the DS estimator over other estimators even including LASSO and OMP. As a reference, the oracle estimator is also plotted as MSE lower bound. It is worth noting that LS-based channel estimation method which cannot robust probed the position of channel dominant coefficients because we considered undermined channel estimation problem. Thus, we can find that even though the SNR up 30dB, the estimation performance just has a little improvement.

## B. MSE performance versus number of training sequence

Next, we compare the performance of the DS and other estimators versus different number of training sequence, which were chosen between 10 and 55. It is seen that, as expected, the performance of all estimators improves with growing number of training sequence. We again observe large performance gains of the DS estimator over other estimators even including LASSO and OMP. It is worth noting that LS-based channel estimation method which cannot robust probed the position of channel dominant coefficients because we considered undermined channel estimation problem. Thus, we can find that even though the number of training sequence increasing from 10 to 55, the estimation performance improved very slowly.

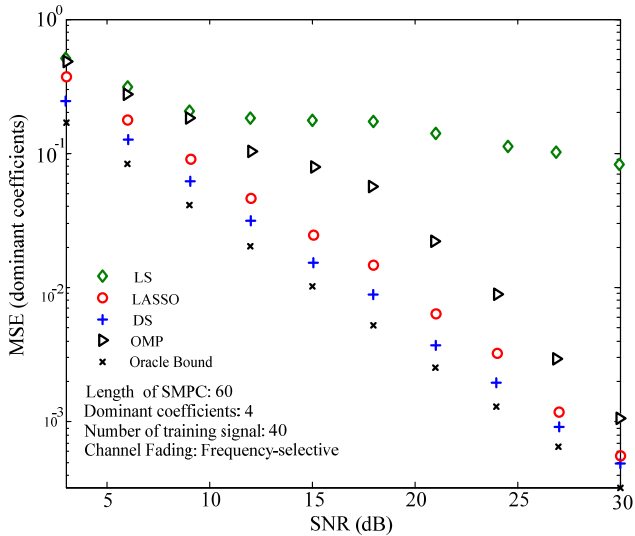

Figure 3. MSE performance of DS and other channel estimation methods

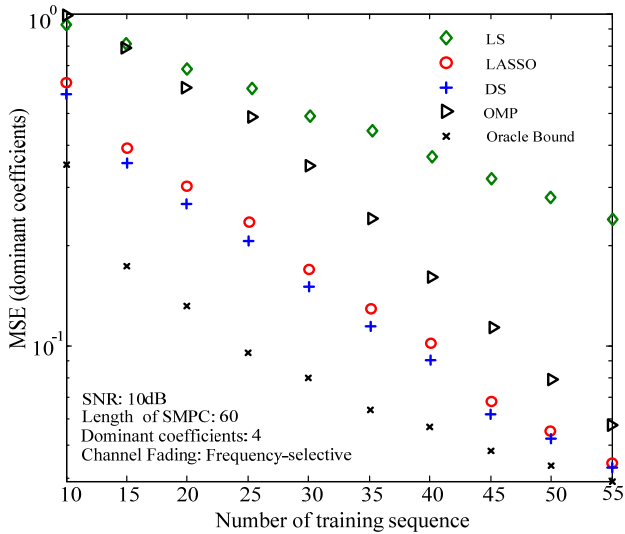

Figure 4. MSE performance of DS and other channel estimation methods versus different number (10~55) of training sequence

## VI. CONCLUSIONS

In this paper, we have introduced a sparse channel estimation technique using DS algorithm which is based on the recently compressed sensing (CS) theory. Various computer simulation results demonstrate that DS algorithm to exploit the sparsity of wireless multipath channel make it possible to reduce the number of sensing training sequence on same performance relative to previous methods. The DS algorithm, hence, can be increase spectral efficiency. According to the development of principal of CS, we conjecture that the estimation performance of the proposed technique can be further improved with SDS algorithm.

### ACKNOWLEDGEMENT

This work is supported in part by the National Natural Science Foundation of China under grant 60772146, the National High Technology Research and Development Program of China (863 Program) under grant 2008AA12Z306, the Key Project of Chinese Ministry of Education under grant 109139 as well as Open Research Foundation of Chongqing Key Laboratory of Signal and Information Processing (CQKLS&IP), Chongqing University of Posts and Telecommunications (CQUPT). And this work is also supported in part by China Scholarship of China Scholarship Council under grant No. 2009607029 and Outstanding Doctor Candidate Training Fund of University of Electronic Science and Technology of China. And this work is also supported in part by Tohoku University Global COE program "Global Education and Research Center for Earth and Planetary Dynamics".